# Auditing Lustre file system

**Sayed Erfan Arefin**
Texas Tech University

**ABSTRACT**
With the increasing demand for data storage and the exponential growth of data, traditional single-server architectures are no longer sufficient to handle the massive amounts of data storage, transfer, and various file system events. As a result, distributed file systems have become a necessity to address the scalability challenges of file systems. One such popular distributed file system is Lustre, which is extensively used in high-performance computing environments. Lustre offers parallel file access, allowing multiple clients to access and store data simultaneously. However, in order to ensure the security and integrity of data, auditing plays a crucial role. Lustre auditing serves as a proof of security and enables the implementation of robust security features such as authentication with Kerberos, mandatory access control with SELinux, isolation, and more. Auditing helps track and monitor file system activities, providing valuable insights into user actions, system events, and potential security breaches. The objective of this project is to explore Lustre auditing using CentOS, a popular Linux distribution, within a Lustre architecture. By implementing Lustre auditing, we aim to enhance the security and reliability of the file system. Additionally, we plan to develop a graphical interface that presents the auditing features in a user-friendly and visually appealing manner. This interface will provide administrators and users with a convenient way to monitor and analyze auditing logs, view access patterns, detect anomalies, and ensure compliance with security policies. By combining the power of Lustre's parallel file system architecture with comprehensive auditing capabilities and an intuitive graphical interface, we aim to provide a robust and user-friendly solution for managing and securing large-scale data storage and access

**INTRODUCTION**
In today's rapidly evolving computer technology landscape, there is a growing need for reliable and high-performance storage systems. Traditional file system models that rely on a single server architecture are no longer capable of meeting the increasing demands of data storage and access. This is where distributed file systems come into play.

Distributed file systems have emerged as a solution to address the limitations of traditional file systems. These systems are designed to distribute data across multiple servers or nodes, allowing for improved scalability, availability, and performance. By leveraging a network-centric approach, distributed file systems can effectively handle large volumes of data and provide efficient access to it.

Unlike single server architectures, where the storage capacity and processing power are limited to a single machine, distributed file systems distribute the data across multiple nodes. This distribution of data allows for parallel processing and load balancing, ensuring that the storage and retrieval of data can be performed in a highly efficient manner.

Furthermore, distributed file systems provide enhanced scalability, allowing organizations to easily scale their storage infrastructure as their data requirements grow. With distributed file systems, additional storage capacity can be seamlessly added by incorporating new nodes into the system, ensuring that the storage capabilities can be expanded without disrupting the overall system performance.

In addition to scalability, distributed file systems offer improved availability. By replicating data across multiple nodes, these systems provide redundancy and fault tolerance. If one node fails or becomes inaccessible, the data can still be accessed from other available nodes, ensuring uninterrupted access to critical information.

Moreover, distributed file systems provide enhanced performance by enabling parallel processing and data access. With data distributed across multiple nodes, tasks can be executed concurrently, resulting in faster data processing and reduced latency. This is particularly beneficial for applications that require high-speed data access and real-time processing.

Overall, distributed file systems have become crucial in modern computer technology due to their ability to provide reliable, scalable, and high-performance storage solutions. These systems are increasingly being adopted in various domains, including cloud computing, big data analytics, scientific research, and high-performance computing, where the efficient management of large volumes of data is essential.

One such distributed file system is Lustre, which stands out as a prominent parallel file system widely used in high-performance computing environments. Lustre addresses the need for scalability and complexity, but it also introduces challenges such as hardware and software vulnerabilities, inconsistent metadata, security threats, and administrative issues [4] [2]. To mitigate these challenges, Lustre auditing plays a vital role by detecting inconsistencies, failures, and administrative errors.

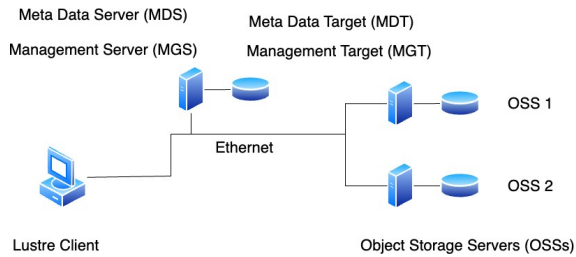

Figure 1. Lustre parallel file system

Figure 2. All the MDT and OSTs

The Lustre system allows users to read, write, and store data using dedicated physical machines that serve specific roles within the system. It enables the scaling of file system performance for handling massive data. The Lustre file system comprises three main components: one or more clients, a metadata service (MDS), and one or more object storage services (OSS), as shown in Figure 1. Each component operates with its own Lustre-specific logic, but they converge at the networking layer. For instance, the MDS handles storage object allocation on the OSS, the OSS provides data storage to clients, and clients interact with the file system by making requests for file system events.

In this project, we gained a comprehensive understanding of the Lustre file system and implemented it using four different virtual machines to represent the three Lustre components. We demonstrated the auditing capabilities of Lustre by integrating a MongoDB database to store and present the auditing information.

By exploring Lustre and its auditing features, we contribute to the overall understanding of distributed file systems and their role in meeting the demands of modern storage requirements. Additionally, the graphical interface provided by the MongoDB integration allows for convenient monitoring and analysis of Lustre auditing data, facilitating the identification of potential issues and ensuring the security and reliability of the file system.

### EXPERIMENTAL SETUP
Settingup the experimental environment with virtual machines and the Lustre file system it self is very important and can be challenging. In this section I discussed how to setup the Virtual Machines and install Lustre on those. Furthermore, the configurations for each VM is provided. I also discussed how to store the changelog in the mongoDB database.

#### Setting up the Virtual Machines
To set up the Lustre system, four virtual machines (VMs) were utilized, each serving as a specific component of the Lustre system. These components include one VM for the client, one VM for the Metadata Server (MDS), and two VMs for the Object Storage Servers (OSS).

The VMs were created using Oracle VirtualBox, a popular virtualization software. Each VM was configured with 2GB of memory and 20GB of disk space, ensuring sufficient resources for running the Lustre system and storing data.

For the operating system, CentOS version 8.3 was chosen. CentOS is a Linux distribution that is known for its stability, reliability, and compatibility with Lustre. The specific kernel version used was 4.18.0-240.1.1.el8_x86_64, which provides necessary features and enhancements for Lustre functionality.

On the MDS (Meta Data Server), one Meta Data Target (MDT) was created. The MDT is responsible for storing the metadata of the Lustre file system, such as file attributes and directory structures. On each of the OSSs (Object Storage Servers), a total of 6 Object Storage Targets (OSTs) were created. The OSTs are responsible for storing the actual data of the Lustre file system, including file contents.

To verify this setup, the command "lfs df -h" was executed on the client. This command provides information about the Lustre file system, including disk usage, available space, and the mapping of OSTs to MDTs. The result of running this command would provide detailed information about the MDT and OST configuration, which can help in understanding the storage setup of the Lustre file system.

The figure 2 represent the output of the "lfs df -h" command, showing the storage configuration and usage statistics of the Lustre file system.

By setting up these four VMs with appropriate specifications and operating systems, the foundation was laid for establishing the Lustre system. Each VM played a vital role in the Lustre architecture, enabling functions such as client-server communication, metadata management, and object storage.

With this setup, the Lustre system was ready to be installed and configured on the respective VMs, facilitating the exploration and utilization of Lustre's features and capabilities.

#### Lustre Installation
For installing, we copied the lustre packages to an HTTP server on the network with the integration into local YUM repositories. We used the following instructions to establish a web server as a YUM repository host for the Lustre packages.

1. At first we created a temporary YUM repository definition using following commands which is used to assist the initial acquisition of Lustre packages.

```
cat >/tmp/lustre-repo.conf <<\_EOF
[lustre-server]
name=lustre-server
baseurl=https://downloads.whamcloud.com/public/lustre
```

```
        /lustre-2.14.0/el8.3.2011/server
# exclude=*debuginfo*
gpgcheck=0

[lustre-client]
name=lustre-client
baseurl=https://downloads.whamcloud.com/public/lustre
        /lustre-2.14.0/el8.3.2011/client
# exclude=*debuginfo*
gpgcheck=0

[e2fsprogs-wc]
name=e2fsprogs-wc
baseurl=https://downloads.whamcloud.com/public/
e2fsprogs/latest/el8
# exclude=*debuginfo*
gpgcheck=0
__EOF
```

2. Next, we used the reposync command (distributed in the yum-utils package) to download mirrors of the Lustre repositories to the manager serve using following commands:

```
mkdir -p /var/www/html/repo
cd /var/www/html/repo
reposync -c /tmp/lustre-repo.conf -n
    --repoid=lustre-server  --repoid=lustre-client
    --repoid=e2fsprogs-wc
```

3. Then, we created the repository metadata with following instructions:

```
yum install createrepo -y
cd /var/www/html/repo
for i in e2fsprogs-wc lustre-client lustre-server; do
(cd $i && createrepo .)
done
```

4. Afterwards, we created a file containing repository definitions for the Lustre packages and stored it in the web server static content directory which makes it easier to distribute to the Lustre servers and clients. The commands we used are as follows:

```
hn=`hostname --fqdn`
cat >/var/www/html/lustre.repo <<__EOF
[lustre-server]
name=lustre-server
baseurl=https://$hn/repo/lustre-server
enabled=0
gpgcheck=0
proxy=_none_
sslverify=0

[lustre-client]
name=lustre-client
baseurl=https://$hn/repo/lustre-client
enabled=0
gpgcheck=0
sslverify=0

[e2fsprogs-wc]
name=e2fsprogs-wc
baseurl=https://$hn/repo/e2fsprogs-wc
enabled=0
gpgcheck=0
sslverify=0
__EOF
```

5. We installed Lustre Server Software by installing e2fsprogs, installing and upgrading the kernel, installing ldiskfs kmod and lustre packages. Next, loaded lustre to the kernel. The commands we used are given below:

```
yum --nogpgcheck --disablerepo=*
    --enablerepo=e2fsprogs-wc \
install e2fsprogs
yum --nogpgcheck --disablerepo=base,extras,updates \
--enablerepo=lustre-server install \
kernel \
kernel-devel \
kernel-headers \
kernel-tools \
kernel-tools-libs \
kernel-tools-libs-devel

reboot

yum --nogpgcheck
    --enablerepo=var_www_html_repo_lustre-server
    install \
kmod-lustre \
kmod-lustre-osd-ldiskfs \
lustre-osd-ldiskfs-mount \
lustre \
modprobe -v lustre
modprobe -v ldiskfs
```

6. Later, we installed the lustre client and for that we have to follow the commands listed in step 5 for upgrading the kernel.Then, we installed the kmod package for lustre client using following commands and loaded lustre to kernel again:

```
yum --nogpgcheck --enablerepo=lustre-client install \
kmod-lustre-client \
Lustre-client
modprobe -v lustre
```

**Lustre Configuration**

The Lustre file system configuration steps should always be performed in the following order:

• Metadata Server

• Object Store Servers

• Client

After the client is configured the lustre file system is usable. SELinux and the firewall was disabled in all the virtual machines. Configuration for all the components are as follows.

*Metadata Server*

The following steps were executed for MDS.

1. The baseline for the MDS is creating a file system named lustre on the server including the metadata target (–mdt) and the management server (–mgs).

```
mkfs.lustre --fsname lustre --mdt --mgs /dev/vg00/mdt
```

2. A mount point was created.

```
mkdir /mdt
```

3. MDS was started using the following command:

```
mount -t lustre /dev/vg00/mdt /mdt
```

*Object Store Servers*

The following steps were executed for each OSS.

1. Mount points were created for the OSTs; this example uses six named ost1 through ost6:

```
mkdir /mnt/ost1

mkdir /mnt/ost6
```

2. The command mkfs.lustre command was used to create the Lustre file systems.

```
mkfs.lustre --fsname lustre --ost --index=0
    --mgsnode=192.168.0.114@tcp0  /dev/vg00/ost1
...
mkfs.lustre --fsname lustre --ost --index=0
    --mgsnode=192.168.0.114@tcp0  /dev/vg00/ost6
```

3. The OSS was started by mounting the OSTs to the corresponding mounting points created in the previous step. The following commands were used for all the mounting points on all the OSSs.

```
mount -t lustre /dev/vg00/ost1 /mnt/ost1

mount -t lustre /dev/vg00/ost6 /mnt/ost6
```

*Client*

The client was setup by mounting the lustre mounting point using the following commands.

1. First, a mounting point was created using the following command.

```
mkdir /mnt/lustre
```

2. The client completed setup by mounting the client mount point using the following command.

```
mount -t lustre 192.168.1.114@tcp0:/lustre /mnt/lustre
```

**FILE SYSTEM AUDITING**

File system auditing plays a crucial role in ensuring data organization and maintaining data integrity within an organization. It provides a means to track and monitor file system activities, allowing for the verification and validation of data stored within the system. By implementing file system auditing, organizations can establish a level of trust and accountability in their data management practices.

There are several key criteria associated with file system auditing:

- Data Modification Tracking: File system auditing tracks any modifications made to files or directories within the system. It records changes such as file creation, modification, deletion, or permission updates. This criterion helps to identify any unauthorized or suspicious activities that may compromise data integrity.

- Access Control Monitoring: File system auditing monitors and logs user access to files and directories. It captures information about who accessed the data, when the access occurred, and the type of action performed. This allows organizations to track and enforce access control policies, ensuring that only authorized individuals can access sensitive information.

- Compliance and Regulatory Requirements: File system auditing helps organizations meet compliance and regulatory requirements by providing an audit trail of file system activities. This criterion ensures that organizations can demonstrate adherence to industry-specific regulations, data protection laws, and internal policies.

- Forensic Investigation: File system auditing facilitates forensic investigation in the event of a security incident or data breach. The audit logs provide valuable information for analyzing and reconstructing the sequence of events leading to the incident, identifying the source of the breach, and implementing appropriate remediation measures.

- Data Integrity Verification: File system auditing verifies the integrity of data stored within the system. By comparing checksums or digital signatures, it ensures that files have not been tampered with or modified in an unauthorized manner. This criterion is particularly important for critical or sensitive data, where maintaining data integrity is paramount.

Overall, file system auditing is a vital component of data management and security. It helps organizations maintain data integrity, enforce access controls, meet compliance requirements, and facilitate forensic investigations. By implementing robust file system auditing practices, organizations can enhance data protection, mitigate risks, and ensure the trustworthiness of their stored data.

**IMPLEMENTATION**

In the Lustre file system, the "changelog" command plays a significant role in facilitating file system auditing. The changelog captures and records essential information that is crucial for auditing purposes. By utilizing the changelog, administrators and security personnel can gain insights into the actions and events occurring within the Lustre file system.

One key aspect of the changelog is its ability to provide comprehensive information about the objects of action. It associates each action with a unique identifier known as the File Identifier (FID). This identifier enables the tracking of specific files or directories that are involved in the audited actions. Additionally, the changelog also captures the name of the targets, which helps identify the specific Lustre components or objects on which the action occurred.

Furthermore, the changelog provides valuable details about the subjects of action. It includes information such as User ID (UID), Group ID (GID), and Network Identifier (NID). This information allows administrators to identify the individuals or entities responsible for the audited actions. By associating the actions with specific users or groups, accountability can be established within the file system.

The timestamp associated with each entry in the changelog is another critical component. It records the time at which each action took place. This timestamp information provides a chronological order of events and enables administrators to trace the sequence of actions within the file system. By analyzing the timestamps, it becomes possible to determine the time of action for auditing and forensic purposes.

The combination of the FIDs, target names, UID/GID, and timestamps in the changelog entries provides a comprehensive audit trail for the Lustre file system. This trail allows for the thorough examination and analysis of file system activities, ensuring data integrity, security, and compliance with regulations. The changelog's ability to capture essential information related to objects, subjects, and time of action makes it a valuable tool for auditing and monitoring Lustre file system operations.

**Prepare for changelog**

In order to prepare our implementation to be able to get changelogs from the MDS virtual machine, the following commands were executed.

1. It is required to register the user to receive changelogs. To register a new changelog user for a device ( example: lustre-MDT0000 ) the following command was executed.

```
mds# lctl --device lustre-MDT0000 changelog_register
output: lustre-MDT0000: Registered changelog userid
    'cl1'
```

2. It is also required to enable all changelog entry types. In order to enable all changelog entry types the following command was executed.

```
mds# lctl set_param
    mdd.lustre-MDT0000.changelog_mask=ALL
output: mdd.seb-MDT0000.changelog_mask=ALL
```

An example of ta changelog entry of "OPEN" type can be seen i nthe following example.

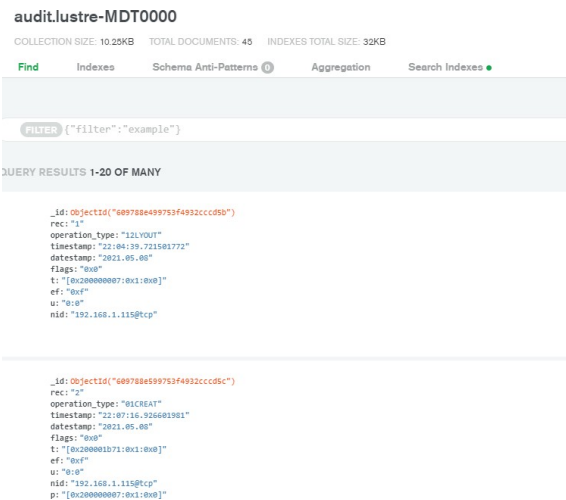

**Figure 3. Documents in mongodb**

```
7 10OPEN 13:38:55.51801258 2017.07.25 0X242
    t=[0x20000401:0x2:0x0] \usepackage{ef=0x7 u=500:500
    nid=10.128.11.159@tcp m=-w-}
```

This command was used ina python program to get changelog ecery 5 seconds. This change logs are then pushed to a mongodb collection. Here the collection is named after the node name. In this case: MDT-0000. The documents are store in the mongodb collection. Afterwards the changelogs are cleared in order to avoid redundancy.

Python 3.8 was used to collect change log information from the MDS. The output of the command was formatted properly and stored in the mongodb. A snapshot of the mongodb collection after running the MDS and executing dummy 100 file creation of 10 MB files from the client can be seen in figure 3. All the possible changelog record type can be observed in Table 1 [1].

**RELATED WORKS**

In their research, Dong D et al conducted a comprehensive performance study of Lustre file systems. One specific aspect they investigated was the LFSCK (Lustre File System Consistency Check) operation, which is responsible for detecting and repairing inconsistencies within the Lustre file system.

During their study, they discovered that in certain scenarios, there can be cascading errors that occur during the LFSCK operation. These cascading errors can propagate and potentially become unrecoverable, leading to a significant problem, particularly in the context of High-Performance Computing (HPC) systems.

The unrecoverable nature of these cascading errors implies that the integrity and reliability of the Lustre file system may be compromised. This finding highlights the importance of addressing and mitigating such issues to ensure the proper

| Value | Description | Value | Description |
|---|---|---|---|
| MARK | Internal recordkeeping | LYOUT | Layout change |
| CREAT | Regular file creation | TRUNC | Regular file truncated |
| MKDIR | Directory creation | SATTR | Attribute change |
| HLINK | Hard link | XATTR | Extended attribute change (setxattr) |
| SLINK | Soft link | HSM | HSM specific event |
| MKNOD | Other file creation | MTIME | MTIME change |
| UNLNK | Regular file removal | CTIME | CTIME change |
| RMDIR | Directory removal | ATIME * | ATIME change |
| RENME | Rename, original | MIGRT | Migration event |
| RNMTO | Rename, final | FLRW | File Level Replication: file initially written |
| NOPEN * | Denied open | RESYNC | File Level Replication: file re-synced |
| CLOSE | Close | GXATR * | Extended attribute access (getxattr) |

**Table 1. Type of records**

functioning and stability of Lustre file systems in HPC environments [3].

Arnab K et al. conducted a study focused on file system monitoring, with specific emphasis on the Lustre parallel file system. They observed that while there are numerous tools available for monitoring desktop file systems, there is a lack of dedicated tools for effectively monitoring Lustre file systems.

To address this gap, the researchers proposed a monitoring tool specifically designed for scalable file systems, with a particular focus on Lustre. The aim of their tool is to provide comprehensive monitoring capabilities that cater to the unique characteristics and requirements of Lustre parallel file systems.

By developing this monitoring tool, Arnab K et al. aimed to enhance the observability and performance analysis of Lustre file systems. This tool would enable system administrators and users to monitor key metrics, track performance bottlenecks, detect anomalies, and gain insights into the overall health and behavior of the Lustre file system.

The development of a dedicated monitoring tool for Lustre file systems is significant as it contributes to the improvement of system management and performance optimization in large-scale parallel computing environments [5].

**CONCLUSION AND FUTURE WORKS**

In this study, the focus was on implementing lustre installation, configuration, and file auditing functions. The lustre file system was set up and configured, allowing for data collection and storage in a mongodb database. The collected data from the lustre file system can be queried and analyzed using various queries.

Moving forward, there are several future directions for this project. One aspect is to enhance the usability of the lustre auditing file system by creating an interactive dashboard using nodejs. This dashboard can provide a user-friendly interface for users to easily access and interpret the collected audit data. It can include features such as data visualization, filtering, and advanced search capabilities, making it more convenient for users to analyze and understand the file system's behavior.

Additionally, the project can explore the application of machine learning techniques to detect security vulnerabilities based on the collected audit data. By utilizing machine learning classifiers, patterns and anomalies in the audit data can be identified, allowing for the detection of potential security threats or vulnerabilities within the lustre file system. This can provide valuable insights for system administrators and help improve the overall security of the file system.

Overall, this project aims to improve the functionality and security of the lustre file system by implementing auditing capabilities, developing a user-friendly dashboard, and leveraging machine learning techniques for vulnerability detection.

**ACKNOWLEDGMENTS**